%% file: main.tex
\documentclass[letterpaper,twocolumn,10pt]{article}
\usepackage{usenix,epsfig,endnotes}

\newcount\Arxiv \Arxiv=1 %

\usepackage{url}
\PassOptionsToPackage{hyphens}{url} %
 \expandafter\def\expandafter\UrlBreaks\expandafter{\UrlBreaks%
   \do\a\do\b\do\c\do\d\do\e\do\f\do\g\do\h\do\i\do\j%
   \do\k\do\l\do\m\do\n\do\o\do\p\do\q\do\r\do\s\do\t%
   \do\u\do\v\do\w\do\x\do\y\do\z\do\A\do\B\do\C\do\D%
   \do\E\do\F\do\G\do\H\do\I\do\J\do\K\do\L\do\M\do\N%
   \do\O\do\P\do\Q\do\R\do\S\do\T\do\U\do\V\do\W\do\X%
   \do\Y\do\Z}

\usepackage{xspace}
\usepackage{xcolor}
\usepackage{tikz}\usetikzlibrary{cd} 
\usepackage{colortbl} %
\usepackage{amsmath}
\usepackage{eurosym}

\usepackage{subcaption}

\usepackage{booktabs} %
\newenvironment{tab}[1]
{%
\begin{tabular}{@{}#1@{}}
\toprule
}
{\bottomrule
\end{tabular}}

\newcount\Comments  \Comments=0 %
\newcommand{\kibitz}[2]{\ifnum\Comments=1\textcolor{#1}{#2}\fi}

\newcommand{\unchecked}{\colorbox{yellow!30}}
\newcommand{\unverifiedblock}{\colorbox{orange!30}}

\begin{document}

\date{}

\title{\Large \bf A Bestiary of Blocking\\ The Motivations and Modes behind Website Unavailability}

\ifnum\Arxiv=0
\author{
{\rm Your N.\ Here}\\
Your Institution
\and
{\rm Second Name}\\
Second Institution
} %
\else
\author{
{\rm Sadia Afroz}\\ {\rm ICSI and UC Berkeley}
\and {\rm Mobin Javed}\\ {\rm LUMS}
\and {\rm Vern Paxson}\\ {\rm ICSI and UC Berkeley}
\and {\rm Shoaib Asif Qazi}\\ {\rm LUMS}
\and {\rm Shaarif Sajid}\\ {\rm LUMS}
\and {\rm Michael Carl Tschantz}\\ {\rm ICSI}
}
\fi

\maketitle

\thispagestyle{empty}

\subsection*{Abstract}

This paper examines different reasons the websites may vary
in their availability by location. Prior
works on availability mostly focus on censorship by nation states.  We
look at three forms of server-side blocking: blocking visitors from
the EU to avoid GDPR compliance, blocking based upon the visitor's
country, and blocking due to security concerns.  We argue that these
and other forms of blocking warrant more research.

\section{Introduction}

We often conceptualize the Internet as one global and shared infrastructure 
comprehensively connecting people from every part of the world.
In practice, however, different users experience different Internets.
The differences in experience can arise for various reasons, such as 
 ISPs creating restricted ``walled gardens'' for their customers, 
 governments censoring access to resources, 
 copyright regulations restricting access to protected content, and 
 web servers blocking unwanted access. 
These partionings of the Internet in terms of the way content is served to 
different end users reflect instances of the 
``balkanization''\footnote{For a critical discussion of the term see Maurer~\cite{maurer14slate}.}
of the Internet into a ``splinternet''.

Currently, a large body of research exists on understanding access restrictions by authoritarian states for censorship 
(e.g.,~\cite{zittrain2003internet,
clayton2006ignoring,
Lowe2007a,
crandall2007conceptdoppler,
park2010empirical,
sfakianakis2011censmon,
syria2011bluecoat,
bamman12first-monday,
verkamp2012inferring,
winters12foci,
khattak2013towards,
nabi2013anatomy,
aryan2013internet,
Dalek2013a,
anon2014towards,
Chaabane2014a,
Marczak2015b,
Ensafi2015b}).
Other forms of access restriction, often controlled by algorithms
running on a website's servers or CDNs, have received much less
empirical exploration.

Herein, we explore forms of that blocking that do not fit into the
quintessential conceptualization of censorship---we leave it to the
reader to decide which count as censorship.
After enumerating various types of blocking (Section~\ref{sec:types}),
we present measurements for three such forms of blocking. 
We show that the number of pages unavailable from three locations in
the EU increased after the EU's General Data Protection Regulation (GDPR)
went into effect (Section~\ref{sec:gdpr}).
We also look for, and find, other blockpages explicitly listing
geography as the reason for blocking (Section~\ref{sec:country}).
Finally, we look for blockpages and practices suggesting security
concerns (Section~\ref{sec:abuse}).

While many of these issues have been discussed, and in some cases measured,
in isolation, we believe this work to be the first to consider the range
the blocking in a systematic manner.  We also discuss the difficulties of
seperating out each form from the others, which is further complicated
by the possiblity of a single block corresponding to more than one form.
Our contributions are, admittedly preliminary:
our list of blocking types is incomplete and tilted toward
location-based blocking; our measurement studies are small-scale.
Nevertheless, we believe there to be value in laying out this space
of research opportunities while highlighting the risks of
claiming to measure only a single phenomenon given the lack of isolation
between the types of blocking we consider.

\paragraph{Prior Work.}
We know of no prior works laying out the space of forms of blocking,
the goal of this work.
However, there are numerous papers looking at various forms of
blocking in isolation.
Thus, rather than have a dedicated prior work section, we will cover
these works where we discuss the form of blocking they cover, mostly in
the next section.

\section{Types of Blocking}
\label{sec:types}

Suppose you run a test to find that a website will load in the US but
not in China.
If the website is politically sensitive, it is not unreasonable to
suspect censorship, but numerous other possibilities exist.

Perhaps the first thing to check is nature of the block: 
was it just a transient network failure?\@
did DNS fail?\@
is there a \textsc{captcha}?\@
a blockpage providing an explanation?\@
a blockpage without an explanation?\@
or an error message?
For example, while \textsc{captcha}s can be annoying, they seem like
an unlikely choice for censorship since they, when working as
designed, merely slow down the accessing of data.
Each of the others seem like stronger indications of censorship,
including, to a lesser extent, even blockpages claiming the cause to
be something else, since censors may mislead.

Another factor to check is what the blocking is targeting.
The blocking could be targeting something other location, such as
the OS or browser used, automated bots loading pages, 
or being logged out of a service.
For example, some websites block
Tor~\cite{khattak16ndss,singh17usenix}.
Keeping these factors and other factors consistent 
across the two locations can help rule them out,
leading to location-based blocking becoming the most
likely explanation.

Questions will still remain about what sort of location is
targeted by the block.
The targeted location might be geographic, such as a campus,
sub-national region, country, or super-national region.
Alternatively, the targeted location might be defined in terms of
network topology: an IP address, an IP address range, a network, or an AS.
One can also ask whether the blocking is whitelisting or
blacklisting.
In \emph{whitelisting}, a website aims to serve its content to only
visitors within its region.
In \emph{blacklisting}, a website aims to exclude certain regions.
Determining the target of the block and its mode of operation can be
tricky.
For example, a large enough blacklist will look like a whitelist, and
blocking enough IP addresses individually will eventually block a whole
range or even a whole country.

Furthermore, given that geographic blocks are likely implemented by
blocking IP address ranges assigned to the targeted country, 
targeting can be considered at multiple
levels from specification to implementation.
Nevertheless, in Section~\ref{sec:country}, we are able to find
country-based blocking with high confidence by finding blockpages that
confess to it.
Censorship could target any of these notions of location, but the
blacklisting of a whole country
(the government's own country) is the most characteristic of
censorship.

Another factor is where in the network the blocking is happening.
The paragon of censorship is a government-operated middlebox at the
national border.
However, other possibilities exist.
The ISP of the client might be doing the blocking
(e.g.,~\cite{crandall2007conceptdoppler}),
or the ISP of the
server hosting the tested website, or the website itself.
A combination of examining how the block operates and additional
measurements can sometimes determine which of these possibilities it
is~\cite{crandall2007conceptdoppler,xu2011internet,aryan2013internet,afroz18arxiv}.

However, even determining where in the network the blocking is
happening does not definitively reveal whether censorship is in
action.
Suppose, you find that the blocking is done by the ISP of your client
in China.
This could be because the government of China ordered the block or
because the ISP is performing some sort of traffic filtering, in
violation of net neutrality, to raise more revenue or cut costs.

Alternatively, suppose you find that the website's server is doing the blocking.
At first, this might seem to be a clear indication that the block is
for some reason other than censorship, such as concerns over abuse.
However, this could still be an instance of China censoring the
website, just in the more roundabout manner of pressuring the
website into blocking visitors from China.
Indeed, Western companies have altered their websites for Chinese
visitors to comply with China's demands~\cite{carsten14reuters}.
Alternately, it could be the server's country doing the censorship by
ordering the blocking of visitors from China as part of economic
sanctions~\cite{bischof15imc,groll18foreign-policy}.

In fact, government orders are behind many sorts of server-side
blocking that might or might not strike the reader as censorship.
A recent example is the passage of SESTA, a US law holding websites
liable for some third-party content facilitating prostitution, which has
lead to a website geo-blocking the US~\cite{brodkin18ars-technica}.
An example we will explore is websites blocking the EU
to avoid GDPR compliance (Section~\ref{sec:gdpr}).
Where to draw the line as to which count as censorship is unclear to us,
but server-side censorship of one form or another is possible.

With this mind, it is clear that censorship is not merely an issue of
where the blocking is happening or who is doing it, but rather one of why the
blocking is happening, that is, upon whose orders.
In some cases, the roles might be switched from what is expected.
For example, arguably, copyright is a form of censorship in which the
copyright holder gets a government to enforce its
claim~\cite{unblockvideosXXweb,paukner13news,albanesius12pcmag}
In theory, a website could pay a government to implement a regional
block to reduce abuse or increase its market share, leading to an odd
form of hybrid government--corporate censorship.

Before concluding that censorship has happened, other possible motivations
behind the block should be considered.
Table~\ref{tbl:reasons} provides a partial list of different forms of blocking.
\begin{table*}
  \caption{Examples of Motivations behind Location-based Blocking.
    Those marked with $\mbox{}^*$ denote location in the 
    network topology instead of geo-location.  
    $\mbox{}^{**}$ denotes cases where we use a non-location-based 
    block due to not finding a location-based one.}
\label{tbl:reasons}
\centering
\begin{tab}{lll}
                      & Server (including CDNs)             & Middlebox (ISPs, governments)\\ %
\midrule
Political censorship  & Bowing to China's demands~\cite{carsten14reuters} & Great Firewall of China (lots)\\
Economic sanctions    & US websites blocking Cuba~\cite{bischof15imc} \& Iran~\cite{groll18foreign-policy} & \\
Third-party liability & Blocking US due to SESTA~\cite{brodkin18ars-technica} &\\ %
Copyright             & YouTube blocking in Germany~\cite{unblockvideosXXweb,paukner13news} & ISPs blocking the Pirate Bay~\cite{albanesius12pcmag}\\
Other compliance      & GDPR (\S\ref{sec:gdpr})           & \\
Security              & Blocking countries assoc.\ w.\ fraud~\cite{burrell2012invisible,afroz18arxiv} (\S\ref{sec:abuse})& \\
Hosting costs         & CDN fees~\cite{afroz18arxiv}  & \\
Revenue               & Price discrimination~\cite{mikians12hotnets,mikians13conext,vissers14hotpets,hannak14imc} & Net-neutrality~\cite{karr18freepress}$^*$\\
Unintentional         & Slash-dotting~\cite{wired04}$^{**}$ & Overloaded rural links \cite{johnson11www,zheleva13dev}\\
\end{tab}
\end{table*}
One possibility is security concerns, such as fraud and abuse, which is
associated with certain
countries~\cite{burrell2012invisible,afroz18arxiv}.
We look at such blocking in Section~\ref{sec:abuse}.
Another possibility is concerns over the costs of serving traffic to
some countries,
which can be seen as a wasted expense for companies not
targeting that market.
This issue  can be  exacerbated by the serving of traffic to  the developing
world  sometimes  being  more  expensive  than  serving  it  to  the
developed world~\cite{afroz18arxiv}.

Also with profits in mind, companies may engage in blocking to
increase revenue, by charging extra fees to access some regions
or by blocking competitors, violations of net neutrality~\cite{karr18freepress}.
While not blocking, some websites have engaged in price
discrimination, which can also negatively affect some visitors based
upon their
location~\cite{mikians12hotnets,mikians13conext,vissers14hotpets,hannak14imc}.

Finally, blocks can be unintentional, for example, from misconfiguration
or failures caused by lacking enough
bandwidth~\cite{wired04,johnson11www,zheleva13dev}.

Above, we mentioned numerous forms of blocking,
but they are not independent of one another.
For example, some serve as
implementation approaches for others.
Measurement studies must take care not to conflate forms of blocking.
The obvious way of doing so is to just ignore the differences.
Less obvious is conflation by attempting identify a
form using proxies for it without making assumptions explicit, such as
assuming that no server-side blocking is censorship.
Developing methods for distinguishing between blocking types
could also aid the blocked users, who currently struggle
to understand what is happening and why~\cite{Gebhart2017a}

\section{GDPR-Motivated Geo-blocking}
\label{sec:gdpr}

The EU's General Data Protection Regulation (GDPR) contains a wide
range of provisions designed to protect the privacy of people who use
online services and to give them more control over their
data~\cite{eu16gdpr}.
Complying with some of the provisions may require a major shift
in how some websites store and process data about their visitors.
(See Lomas~\cite{lomas18tc} for an overview.)
For example, in general, visitors have the right to access,
correct, and delete data about themselves.
Implementing these abilities
can create an implementational headache given
systems engineered to use and store data indiscriminately.
Furthermore, getting it correct is high stakes, with
fines of
\EUR{20M} or 4\% of a company's global annual revenue.
Given the uncertainty and stakes, some websites have decided to exit the
European market, at least for the time
being~\cite{hern18guardian,hill18register}.
To partly quantify this effect,
we analyze the differences in availability 
of a convenience sample of websites before and
after GDPR went into effect.

\ifnum\Arxiv=0
From Afroz~et~al.~\cite{afroz18arxiv}, we acquired a data set showing
the availability of 7081~websites, which they collected to study a
different facet of server-side blocking.
These websites form the union of various Alexa top 500 lists: the
global list, the lists for ten countries, and the lists for nine
categories of websites.
For each URL, they measured it from three locations in the EU via a
VPN: London, United Kingdom; Sofia, Bulgaria; and Frankfurt,
Germany.
From each location, we use one of their measurements of the
URLs before GDPR came into effect.
The crawler used in these measurements 
attempted DNS resolution for each URL, logging any errors.
For those that resolved, it used Python's Requests package to
request the webpage, logging errors, status codes, and content.
See their work for details~\cite{afroz18arxiv}.

After GDPR came into effect, we re-used the measurement infrastructure of
Afroz~et~al.\@ to take a second measurement of each URL.
We took the data and analyzed it for changes in website availability.
\else
From our prior work~\cite{afroz18arxiv}, we re-used a data set showing
the availability of 7081~websites, which we collected to study a
different facet of server-side blocking.
These websites form the union of various Alexa top 500 lists: the
global list, the lists for ten countries, and the lists for nine
categories of websites.
For each URL, we measured it from three locations in the EU via a
VPN: London, United Kingdom; Sofia, Bulgaria; and Frankfurt,
Germany.
From each location, we use one of the measurements of the
URLs before GDPR came into effect.
The crawler used in these measurements 
attempted DNS resolution for each URL, logging any errors.
For those that resolved, it used Python's Requests package to
request the webpage, logging errors, status codes, and content.
See our prior work for details~\cite{afroz18arxiv}.

After GDPR came into effect, we re-used our measurement infrastructure of
to take a second measurement of each URL.
We took the data and analyzed it for changes in website availability.
\fi

\input{gdpr-ex.tex}
Using error logs and status codes, we found 74~websites that,
for all three European locations, sent an HTTP status code of
200~\emph{OK} when accessed before %
May~25 and non-200 status after May~25, 2018.
Out of the 74~websites, 40 responded with a 403~\emph{Forbidden} status code
and a block page explicitly mentioning ``Blocked due to GDPR'' (Table~\ref{tab:gdpr}). 
Seven websites used the 
HTTP status code 451~\emph{Unavailable For Legal Reasons} (Table~\ref{tab:gdpr-451}), 
the code named for a novel on censorship~\cite{bradbury53book}.
All of 47 of these websites with explicit blockpages are local news websites,
incidentally, a plausible target of censorship as well.
One website, \url{brownells.com}, asks users to visit their EU website, \url{www.brownells.eu}.
The remaining 27~websites do not
provide any explicit blockpages and use rather vague status codes and connection errors. 
For example, the online gaming website \url{addictinggames.com} returns an empty page with a 404~\emph{Page Not Found} status code,
the math tutoring website \url{webmath.com} refuses the TCP connection, and  
the reality website \url{99acres.com} responded with a 412~\emph{Precondition Failed}
error code. 
\input{gdpr-451.tex}

Providing context to our findings, looking online, we found services
aiming to make it easy to block all EU
visitors~\cite{parrilla18gdpr,cimpanu18bleeping-computer}.

\section{Country-based Blocking}
\label{sec:country}

Cloudflare, a CDN, is an interesting subject of study, not only because
it hosts many websites, but because it provides more information than
many explaining why it blocks, on the behalf of the website owner, certain
requests.
In this section and the next, we analyze Cloudflare's block notices
looking for country-based and then security-motivated blocking.
We emphasize that we selected Cloudflare not because we believe it to
engage in such blocks any more than any other host, nor because we
believe it should be singled out for criticism. 
Rather, we selected Cloudflare since prior work has found it blocking Tor based
on abuse~\cite{khattak16ndss} and because of the information that
Cloudflare provides about blocks, a commendable feature.

We start by finding sites hosted by Cloudflare.
We resolved the top 1M Alexa domains, and
identified those hosted by Cloudflare by performing \textit{whois} lookups on
the resolved IP addresses and keeping those containing ``Cloudflare'' in the AS name.  
We identified 85,421 Cloudflare-hosted URLs in this fashion.

Next, for each of these Cloudflare-hosted URLs, we retrieved the website from five vantage
points: Pakistan (home network), Scotland (VPN), South Africa (cloud), Ukraine (VPN),
and US (institutional network).
The US crawler experienced
failure, limiting its collection 77,935 URLs.

We then classified the responses.
By examining the blockpages themselves and Cloudflare's
documentation~\cite{gonlag18cloudflare,cloudflare18errors}, we
determined that a response with an HTTP status code of 403 and a body
that referenced Cloudflare's own error code~1009 indicates blocking by
country.
Table~\ref{tbl:cloudflare-blockpages} shows a breakdown of the responses by both this
code and others, some of which will be relevant to the next section.

\begin{figure}
\centering                                                                      
\includegraphics[width=0.5\textwidth]{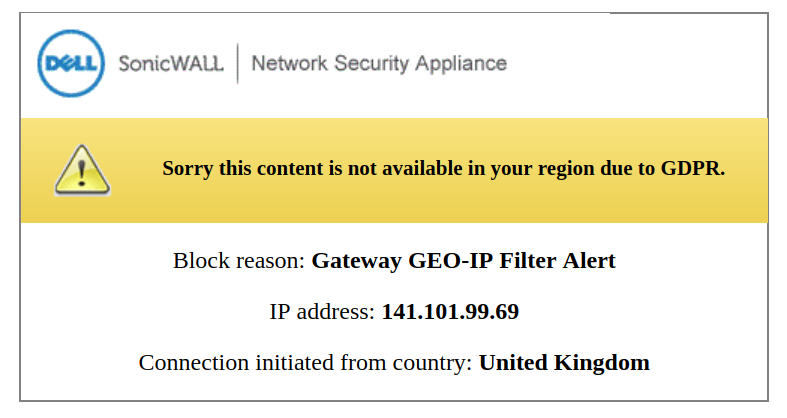}
\caption{SonicWall Blockpage Showing the Motivation to be GDPR avoidance.}
\label{fig:sonicwall}                                            
\end{figure}

We found 524~websites using country-based blocking by Cloudflare under
error code~1009.  We also found one website using country-based blocking by a
different service provider, Dell's SonicWall, which the website \url{motorcar.com} used in addition to
Cloudflare.
The error message for SonicWall's 403 blockpage says ``Sorry this content is not
available in your country due to GDPR'', shown in Figure~\ref{fig:sonicwall},
proving to be instance of both GDPR and country-based blocking.

\begin{figure}
\centering
\includegraphics[width=0.5\textwidth]{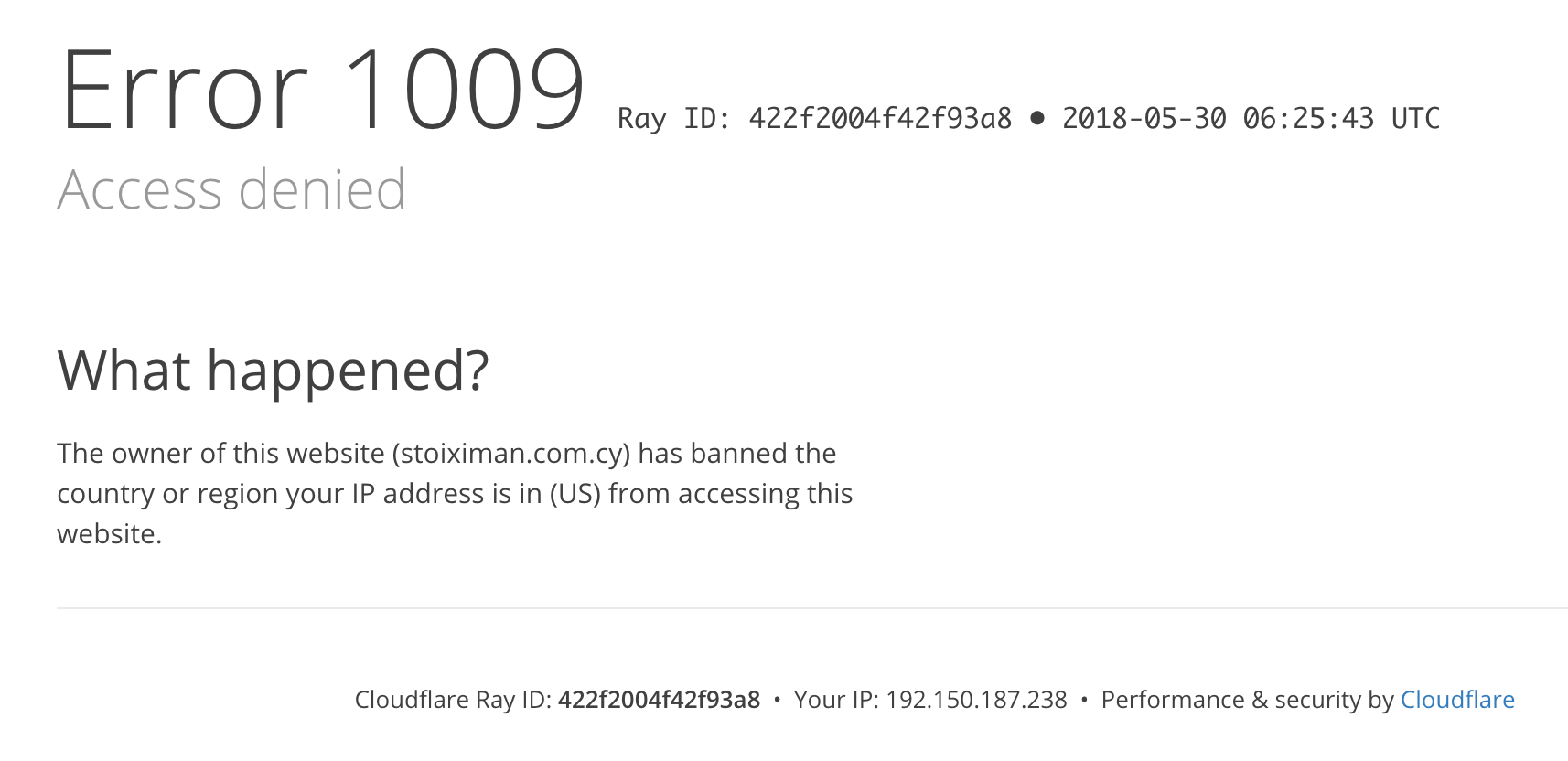}
\caption{Cloudflare Blockpage with a 1009 Error Code}
\label{fig:cf1009}
\end{figure}

Interestingly, 32 websites were country blocked in the US, with 21
giving a Cloudflare 1009~error.
We manually checked all 32
and found five that would load, including one that was give a
1009~error to the crawler.
A different website with a 1009~error was \url{aquapro.biz}.
It was
blocking countries and manually unavailable in the US, but
consistently misidentified our country in an inconsistent manner,
seemingly to assign us other countries at random.
The remaining 1009~errors explicitly blocked the US.
(For an example, see Figure~\ref{fig:cf1009}.)

\begin{table*}
\caption{Blockpage types for 85,421 Cloudflare-hosted domains from various vantage
points.}
\label{tbl:cloudflare-blockpages}
\centering\small
\begin{tab}{rlrrrrr}
 \multicolumn{2}{@{}l}{Blocktypes / Vantage point} & Ukraine  & Scotland & Pakistan & South 
    Africa & USA \\
Status & Description        & (VPN)    & (VPN)  & (Home) & (Cloud) &
    (Inst.) \\
\midrule
\multicolumn{2}{@{}l}{\textbf{No HTTP Response}}    \\
n/a & Timed out & 579 & 542 & 607  & 577 & 540 \\
n/a & DNS error & 45 & 112 & 4096 & 4 & 66 \\
n/a & Other connection errors & 147 & 959 & 132 & 70 & 525 \\[1ex]
\multicolumn{2}{@{}l}{\textbf{Geo-blocking totals}}   & \textbf{313} & \textbf{175} & \textbf{178} & \textbf{103} & \textbf{32} \\
403 & Cloudflare: country or region blocked (1009) & 257 & 161 & 162 & 88 & 21 \\
403 & SonicWall Geo-IP filter                      & 1 & 1 & 0 & 1 & 0 \\
403 & Other blockpage mentioning geo-blocking      & 40 & 11 & 3 & 3 & 0 \\
200 & Other blockpage mentioning geo-blocking      & 15 & 1 & 13 & 11 & 10 \\
451 & Unavailable For Legal Reasons                & 0 & 1 & 0 & 0 & 1 \\[1ex]
\multicolumn{2}{@{}l}{\textbf{Abuse-blocking totals}}       & \textbf{3431} & \textbf{1417} & \textbf{1874} & \textbf{1537} & \textbf{1255} \\
403 & Cloudflare: IP Blocked (1006, 1007, 1008)   & 23 & 5 & 6 & 5 & 1 \\
200 & Cloudflare: IP Blocked (1006, 1007, 1008)   & 2 & 2 & 2 & 2 & 1 \\
503 & Cloudflare: Browser Verification            & 1519 & 1091 & 1111 & 1124 & 985 \\
200 & Cloudflare: Browser Verification            & 2 & 2 & 3 & 2 & 3 \\
403 & Cloudflare: CAPTCHA Challenge        & 1874 & 309 & 746 & 395 & 257 \\
403 & OctoNet HTTP filter: VPN / TOR Block & 3 & 0 & 0 & 0 & 0 \\[1ex]
\multicolumn{2}{@{}l}{\textbf{Misconfigurations totals}}   & \textbf{8}  & \textbf{8} & \textbf{6} & \textbf{9} & \textbf{8}  \\
403 & Cloudflare: DNS points to invalid IP (1000, 1002) & 8 & 8 & 6 & 9 & 8 \\
\end{tab}
\end{table*}

While the rate of country blocks varied from country to country,
this comparison is complicated by the fact that different countries
had different success rates at getting any response from the server.
For example, Pakistan had an abnormally high rate of DNS errors,
possibly due to network failures or censorship.
This difference might hide a much higher rate of blocks in Pakistan
than in Scotland.
Alternatively, Scotland using a VPN and Pakistan using a home network
might hide the difference.
However, Scotland and Ukraine can be compared on a fairly even basis
for both of these factors.
For them, we see a large difference with
Ukraine receiving more blocks.

\section{Security-motivated Blocking}
\label{sec:abuse}

Using the same data set, we also looked for security-motivated blocking.
While we recognize that some country-based blocks may have security as
its motivation, we exclude those that we found to explicitly claiming
to be country-based blocks to focus on those not yet discussed.
We look at other types of blocking that could have been
motivated by security concerns, while admitting that we cannot
be sure of the real motivations behind a block.

Again, Cloudflare's documentation helped us know where to
look~\cite{gonlag18cloudflare,cloudflare18errors}.
The most indicative error code of security concerns is 1010, described
as ``bad browser'', which happens when ``The source of the request was
not legitimate or the request itself was
malicious''~\cite{gonlag18cloudflare}.
Cloudflare also uses error code~1012 to deny access 
``based on malicious activity detected from your computer or your
network (ip\_address)''~\cite{cloudflare17web}.
We also included error codes indicating an IP block, although those
could be used for non-security reasons.

\begin{figure}
\centering
\includegraphics[width=0.5\textwidth]{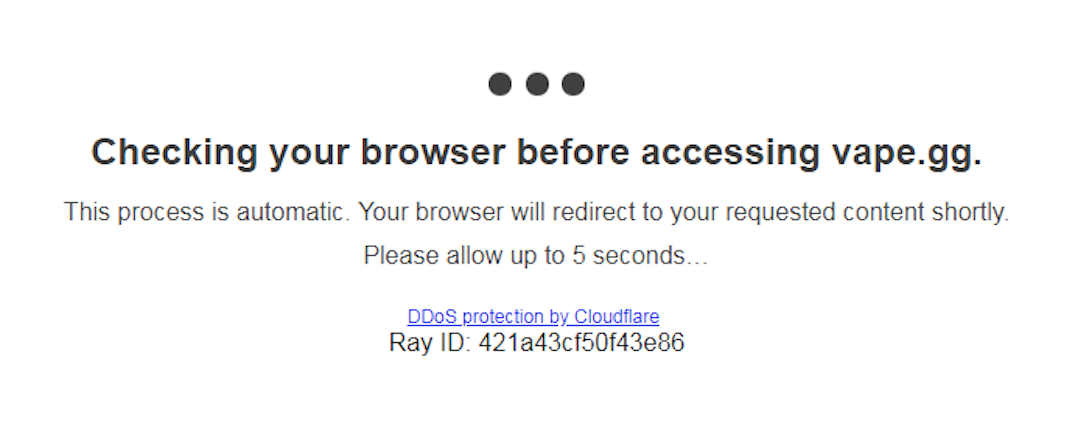}
\caption{Cloudflare's Browser Challenge in Action}
\label{fig:browser-challenge}
\end{figure}

Finally, we looked at restrictions short of outright blocking.
Namely, Cloudflare will sometimes make users solve a \textsc{captcha}
before showing them the page.
Cloudflare will also use a ``browser challenge'' on visitors it
suspects of being a bot,
which we looked for (see Figure~\ref{fig:browser-challenge}).

Table~\ref{tbl:cloudflare-blockpages} shows how common each of these, and other,
outcomes are.
As with country-based blocking, comparing across countries is
confounded.
Looking again at the well matched pair of Scotland and Ukraine,
we see a large difference, with
Ukraine receiving more blocks, \textsc{captcha}s, browser
verifications.
Between the two, only Ukraine was accused of being a VPN or Tor despite both
using the same VPN provider.
The VPN/Tor blockpages came not from Cloudflare, but from OctoNet HTTP
filter.

\section{Conclusions}

We have laid out a space of blocking that includes, but also exceeds,
what we normally think of as censorship.
We looked at three such forms of blocking in some detail.
One of them, country-based blocking, is directly tied to location.
It is more of an approach for implementing blocking than a
motivation for blocking,
raising the question of why the blocking is happening.
The other two forms we measured are more like motivations than
implementation approaches.
The first, avoiding GDPR compliance, is, however, directly related to
location in that the websites are blocking visitors from the EU for
being from the EU.
The third, security-motivated blocks, differs in that it does not need
to be implemented using locations.
However, we do find large differences in how
common security-based blocks are from one location to the next, even
when using the same VPN service to send requests from each location.
While we studied three forms of blocking, they were
far from independent of one another.
While each of our studies were small scale, we hope they stimulate
further research on these issues.

Deciding exactly which of these count as \emph{censorship}
is politically fraught, and we will not attempt to do so.
We do take the stance that research should cover all forms of blocking
that can adversely affect some users, particularly, when those effects
are concentrated on people from certain countries.
Furthermore, we believe the chilling
effects on website availability of even well-intentioned laws to be an
interesting subject of measurement.
While we may wish for a world with both the robust privacy protections
of the GDPR and
an Internet free from balkanization, currently, a tradeoff is evident.
The blocking of EU visitors precipitated by a privacy law may even
have an outsized effect on Tor given the outsized number of Tor exits
in the EU.
This serves as an example of how the forms of blocking do not just have
interdependencies between themselves but also with privacy.
The presence of these interdependencies should be kept in mind when
attempting to measure censorship to avoid false positives.

\ifnum\Arxiv=1
\paragraph{Acknowledges}
We thank David Fifield for allowing us to use his code for
categorizing block pages.
We gratefully acknowledge funding support from the National Science Foundation (Grants 1518918 and 1651857) and UC~Berkeley's Center for Long-Term Cybersecurity. The opinions in this paper are those of the authors and do not necessarily reflect the opinions of any funding sponsor or the United States Government.
\fi

{\footnotesize \bibliographystyle{acm}
\bibliography{censor}}

\end{document}

%% file: gdpr-ex.tex
\begin{table}
\caption{Websites explicitly mentioning GDPR as motativation for blocking. The \emph{Before} column shows the status for all the vantage points. DE represents DNS Error.}
\label{tab:gdpr} 
\small
\centering
\setlength{\tabcolsep}{2pt}
\begin{tab}{lr@{\hspace{20pt}}rrrr}
&     Before  & \multicolumn{4}{c}{After}\\
\cmidrule(l){3-6} 
URL/Country & & US & BGR & GBR & DEU\\
\midrule
bismarcktribune.com  & 200  & 200  & \unchecked{DE}  & \unverifiedblock{403}  & \unverifiedblock{403} \\ 
collegian.psu.edu  & 200  & 200  & \unverifiedblock{403}  & \unverifiedblock{403}  & \unverifiedblock{403} \\ 
dailynebraskan.com  & 200  & 200  & \unchecked{DE}  & \unverifiedblock{403}  & \unverifiedblock{403} \\ 
dailyprogress.com  & 200  & 200  & \unverifiedblock{403}  & \unverifiedblock{403}  & \unverifiedblock{403} \\ 
fredericknewspost.com  & 200  & 200  & \unverifiedblock{403}  & \unverifiedblock{403}  & \unverifiedblock{403} \\ 
fredericksburg.com  & 200  & 200  & \unverifiedblock{403}  & \unchecked{DE}  & \unverifiedblock{403} \\ 
globegazette.com  & 200  & 200  & \unverifiedblock{403}  & \unverifiedblock{403}  & \unverifiedblock{403} \\ 
greensboro.com  & 200  & 200  & \unverifiedblock{403}  & \unverifiedblock{403}  & \unverifiedblock{403} \\ 
gwinnettdailypost.com  & 200  & 200  & \unverifiedblock{403}  & \unverifiedblock{403}  & \unverifiedblock{403} \\ 
havasunews.com  & 200  & 200  & \unverifiedblock{403}  & \unverifiedblock{403}  & \unverifiedblock{403} \\ 
heraldtimesonline.com  & 200  & 200  & \unverifiedblock{403}  & \unverifiedblock{403}  & \unverifiedblock{403} \\ 
host.madison.com/wsj  & 200  & 200  & \unverifiedblock{403}  & \unverifiedblock{403}  & \unverifiedblock{403} \\ 
journalnow.com  & 200  & 200  & \unverifiedblock{403}  & \unverifiedblock{403}  & \unverifiedblock{403} \\ 
journalstar.com  & 200  & 200  & \unverifiedblock{403}  & \unverifiedblock{403}  & \unverifiedblock{403} \\ 
journaltimes.com  & 200  & 200  & \unchecked{DE}  & \unverifiedblock{403}  & \unverifiedblock{403} \\ 
lacrossetribune.com  & 200  & 200  & \unverifiedblock{403}  & \unverifiedblock{403}  & \unverifiedblock{403} \\ 
lancasteronline.com  & 200  & 200  & \unverifiedblock{403}  & \unverifiedblock{403}  & \unverifiedblock{403} \\ 
napavalleyregister.com  & 200  & 200  & \unchecked{DE}  & \unverifiedblock{403}  & \unverifiedblock{403} \\ 
nwitimes.com  & 200  & 200  & \unchecked{DE}  & \unverifiedblock{403}  & \unverifiedblock{403} \\ 
omaha.com  & 200  & 200  & \unverifiedblock{403}  & \unverifiedblock{403}  & \unverifiedblock{403} \\ 
pantagraph.com  & 200  & 200  & \unverifiedblock{403}  & \unverifiedblock{403}  & \unverifiedblock{403} \\ 
pilotonline.com  & 200  & 200  & \unverifiedblock{403}  & \unverifiedblock{403}  & \unverifiedblock{403} \\ 
postandcourier.com  & 200  & 200  & \unverifiedblock{403}  & \unverifiedblock{403}  & \unverifiedblock{403} \\ 
postbulletin.com  & 200  & 200  & \unverifiedblock{403}  & \unverifiedblock{403}  & \unverifiedblock{403} \\ 
pressofatlanticcity.com  & 200  & 200  & \unchecked{DE}  & \unverifiedblock{403}  & \unverifiedblock{403} \\ 
qctimes.com  & 200  & 200  & \unverifiedblock{403}  & \unverifiedblock{403}  & \unverifiedblock{403} \\ 
rapidcityjournal.com  & 200  & 200  & \unverifiedblock{403}  & \unverifiedblock{403}  & \unverifiedblock{403} \\ 
richmond.com  & 200  & 200  & \unverifiedblock{403}  & \unverifiedblock{403}  & \unverifiedblock{403} \\ 
roanoke.com  & 200  & 200  & \unverifiedblock{403}  & \unverifiedblock{403}  & \unverifiedblock{403} \\ 
santafenewmexican.com  & 200  & 200  & \unverifiedblock{403}  & \unverifiedblock{403}  & \unverifiedblock{403} \\ 
southbendtribune.com  & 200  & 200  & \unverifiedblock{403}  & \unverifiedblock{403}  & \unverifiedblock{403} \\ 
stltoday.com  & 200  & 200  & \unverifiedblock{403}  & \unverifiedblock{403}  & \unverifiedblock{403} \\ 
theadvocate.com  & 200  & 200  & \unverifiedblock{403}  & \unverifiedblock{403}  & \unverifiedblock{403} \\ 
trib.com  & 200  & 200  & \unverifiedblock{403}  & \unverifiedblock{403}  & \unverifiedblock{403} \\ 
tucson.com  & 200  & 200  & \unverifiedblock{403}  & \unverifiedblock{403}  & \unverifiedblock{403} \\ 
wacotrib.com  & 200  & 200  & \unverifiedblock{403}  & \unverifiedblock{403}  & \unverifiedblock{403} \\ 
wcfcourier.com  & 200  & 200  & \unchecked{DE}  & \unverifiedblock{403}  & \unverifiedblock{403} \\ 
wvgazettemail.com  & 200  & 200  & \unverifiedblock{403}  & \unverifiedblock{403}  & \unverifiedblock{403} \\ 
yakimaherald.com  & 200  & 200  & \unverifiedblock{403}  & \unverifiedblock{403}  & \unverifiedblock{403} \\ 
\end{tab} 
\end{table}

%% file: gdpr-451.tex
\begin{table}
\caption{Websites mentioning ``Blocked for legal reasons''.  The \emph{Before} column shows the status for all the vantage points. DE represents DNS Error.}
\label{tab:gdpr-451} 
\centering
\setlength{\tabcolsep}{2pt}
\begin{tab}{lr@{\hspace{15pt}}rrrr}
&     Before  & \multicolumn{4}{c}{After}\\
\cmidrule(lr){3-6}
URL/Country & & US & BGR & GBR & DEU\\                                          
\midrule 
ctpost.com  & 200  & 200  & \unverifiedblock{451}  & \unverifiedblock{451}  & \unverifiedblock{451} \\ 
greenwichtime.com  & 200  & 200  & \unverifiedblock{451}  & \unverifiedblock{451}  & \unverifiedblock{451} \\ 
lmtonline.com  & 200  & 200  & \unchecked{DE}  & \unverifiedblock{451}  & \unverifiedblock{451} \\ 
newstimes.com  & 200  & 200  & \unverifiedblock{451}  & \unverifiedblock{451}  & \unverifiedblock{451} \\ 
nhregister.com  & 200  & 200  & \unverifiedblock{451}  & \unverifiedblock{451}  & \unverifiedblock{451} \\ 
seattlepi.com  & 200  & 200  & \unverifiedblock{451}  & \unverifiedblock{451}  & \unverifiedblock{451} \\ 
stamfordadvocate.com  & 200  & 200  & \unverifiedblock{451}  & \unverifiedblock{451}  & \unverifiedblock{451} \\ 
\end{tab} 
\end{table}

%% file: main.bbl
\begin{thebibliography}{10}

\bibitem{afroz18arxiv}
{\sc Afroz, S., Tschantz, M.~C., Sajid, S., Qazi, S.~A., Javed, M., and Paxson,
  V.}
\newblock Exploring server-side blocking of regions.
\newblock {\em ArXiv 1805.11606\/} (May 2018).

\bibitem{albanesius12pcmag}
{\sc Albanesius, C.}
\newblock {U.K.} high court orders {ISP}s to block the {P}irate {B}ay.
\newblock {\em PC Mag\/} (Apr. 2012).
\newblock \url{https://www.pcmag.com/article2/0,2817,2403749,00.asp}.

\bibitem{anon2014towards}
{\sc Anonymous}.
\newblock Towards a comprehensive picture of the {Great Firewall's} {DNS}
  censorship.
\newblock In {\em Free and Open Communications on the Internet\/} (2014),
  USENIX.

\bibitem{aryan2013internet}
{\sc Aryan, S., Aryan, H., and Halderman, J.~A.}
\newblock {Internet} censorship in {Iran}: A first look.
\newblock {\em Free and Open Communications on the Internet, Washington, DC,
  USA\/} (2013).

\bibitem{bamman12first-monday}
{\sc Bamman, D., O'Connor, B., and Smith, N.~A.}
\newblock Censorship and deletion practices in chinese social media.
\newblock {\em First Monday 17}, 3 (Mar. 2012).

\bibitem{bischof15imc}
{\sc Bischof, Z.~S., Rula, J.~P., and Bustamante, F.~E.}
\newblock In and out of cuba: Characterizing cuba's connectivity.
\newblock In {\em Proceedings of the 2015 Internet Measurement Conference\/}
  (New York, NY, USA, 2015), IMC '15, ACM, pp.~487--493.

\bibitem{bradbury53book}
{\sc Bradbury, R.}
\newblock {\em Fahrenheit 451}.
\newblock Ballantine Books), 1953.

\bibitem{brodkin18ars-technica}
{\sc Brodkin, J.}
\newblock ``erotic review'' blocks {US} {I}nternet users to prepare for
  government crackdown.
\newblock {\em Ars Technica\/} (Apr. 2018).
\newblock
  \url{https://arstechnica.com/tech-policy/2018/04/erotic-review-blocks-us-internet-users-to-prepare-for-government-crackdown/}.

\bibitem{burrell2012invisible}
{\sc Burrell, J.}
\newblock {\em Invisible users: Youth in the {I}nternet caf{\'e}s of urban
  {G}hana}.
\newblock Mit Press, 2012.

\bibitem{carsten14reuters}
{\sc Carsten, P.}
\newblock Microsoft denies global censorship of {C}hina-related searches.
\newblock {\em Reuters\/} (Feb. 2014).
\newblock
  \url{https://www.reuters.com/article/us-microsoft-bing-censorship-idUSBREA1B0CP20140212}.

\bibitem{Chaabane2014a}
{\sc Chaabane, A., Chen, T., Cunche, M., Cristofaro, E.~D., Friedman, A., and
  Kaafar, M.~A.}
\newblock Censorship in the wild: Analyzing {Internet} filtering in {Syria}.
\newblock In {\em Internet Measurement Conference\/} (2014), ACM.

\bibitem{cimpanu18bleeping-computer}
{\sc Cimpanu, C.}
\newblock New service blocks {EU} users so companies can save thousands on
  {GDPR} compliance.
\newblock {\em Bleeping Computer\/} (May 2018).
\newblock
  \url{https://www.bleepingcomputer.com/news/security/new-service-blocks-eu-users-so-companies-can-save-thousands-on-gdpr-compliance/}.

\bibitem{syria2011bluecoat}
{\sc {Citizen Lab}}.
\newblock Behind {Blue Coat}: Investigations of commercial filtering in {Syria}
  and {Burma}.
\newblock \url{https://citizenlab.org/2011/11/behind-blue-coat/}, Nov. 2011.

\bibitem{clayton2006ignoring}
{\sc Clayton, R., Murdoch, S.~J., and Watson, R. N.~M.}
\newblock Ignoring the {Great Firewall of China}.
\newblock In {\em Privacy Enhancing Technologies\/} (Cambridge, England, 2006),
  Springer, pp.~20--35.

\bibitem{cloudflare18errors}
{\sc {Cloudflare}}.
\newblock Error pages, troubleshooting, cloudflare support.
\newblock
  \url{https://support.cloudflare.com/hc/en-us/sections/200820298-Error-Pages}.
\newblock Accessed May~29, 2018.

\bibitem{cloudflare17web}
{\sc {Cloudflare Damon}}.
\newblock 1012 error: Access denied.
\newblock Cloudflare support page:
  \url{https://support.cloudflare.com/hc/en-us/articles/200171986-1012-Error-Access-Denied},
  Jan. 2017.

\bibitem{crandall2007conceptdoppler}
{\sc Crandall, J.~R., Zinn, D., Byrd, M., Barr, E., and East, R.}
\newblock {ConceptDoppler}: A weather tracker for {Internet} censorship.
\newblock In {\em Computer and Communications Security\/} (Alexandria, VA, USA,
  2007), ACM, pp.~352--365.

\bibitem{Dalek2013a}
{\sc Dalek, J., Haselton, B., Noman, H., Senft, A., Crete-Nishihata, M., Gill,
  P., and Deibert, R.~J.}
\newblock A method for identifying and confirming the use of {URL} filtering
  products for censorship.
\newblock In {\em Internet Measurement Conference\/} (2013), ACM.

\bibitem{Ensafi2015b}
{\sc Ensafi, R., Fifield, D., Winter, P., Feamster, N., Weaver, N., and Paxson,
  V.}
\newblock Examining how the {Great Firewall} discovers hidden circumvention
  servers.
\newblock In {\em Internet Measurement Conference\/} (New York, NY, USA, 2015),
  ACM, pp.~445--458.

\bibitem{eu16gdpr}
{\sc {European Commission}}.
\newblock General data protection regulation ({GDPR}).
\newblock Regulation (EU) 2016/679, L119, May 2016.

\bibitem{Gebhart2017a}
{\sc Gebhart, G., Author, A., and Kohno, T.}
\newblock Internet censorship in {Thailand}: User practices and potential
  threats.
\newblock In {\em European Symposium on Security \& Privacy\/} (2017), IEEE.

\bibitem{gonlag18cloudflare}
{\sc Gonlag, M.}
\newblock What are the types of threats?
\newblock Cloudflare Support page:
  \url{https://support.cloudflare.com/hc/en-us/articles/204191238-What-are-the-types-of-Threats},
  May 2018.
\newblock Accessed May~29, 2018.

\bibitem{groll18foreign-policy}
{\sc Groll, E.}
\newblock How {W}ashington helps {T}ehran control the {I}nternet.
\newblock {\em Foreign Policy\/} (Jan. 2018).
\newblock
  \url{http://foreignpolicy.com/2018/01/04/how-washington-helps-tehran-control-the-internet/}.

\bibitem{hannak14imc}
{\sc Hannak, A., Soeller, G., Lazer, D., Mislove, A., and Wilson, C.}
\newblock Measuring price discrimination and steering on e-commerce web sites.
\newblock In {\em Proceedings of the 2014 Conference on Internet Measurement
  Conference\/} (New York, NY, USA, 2014), ACM, pp.~305--318.

\bibitem{hern18guardian}
{\sc Hern, A., and Waterson, J.}
\newblock Sites block users, shut down activities and flood inboxes as {GDPR}
  rules loom.
\newblock {\em The Guardian\/} (May 2018).
\newblock
  \url{https://www.theguardian.com/technology/2018/may/24/sites-block-eu-users-before-gdpr-takes-effect}.

\bibitem{hill18register}
{\sc Hill, R.}
\newblock {US} websites block netizens in {E}urope: {W}hy are they ghosting
  {EU}? it's not you, it's {GDPR}.
\newblock {\em The Register\/} (May 2018).
\newblock
  \url{https://www.theregister.co.uk/2018/05/25/tronc_chicago_tribune_la_times_gdpr_lock_out_eu_users/}.

\bibitem{johnson11www}
{\sc Johnson, D.~L., Pejovic, V., Belding, E.~M., and van Stam, G.}
\newblock Traffic characterization and internet usage in rural africa.
\newblock In {\em Proceedings of the 20th International Conference Companion on
  World Wide Web\/} (New York, NY, USA, 2011), WWW '11, ACM, pp.~493--502.

\bibitem{karr18freepress}
{\sc Karr, T.}
\newblock Net neutrality violations: A brief history.
\newblock {\em Free Press\/} (Jan. 2018).
\newblock
  \url{https://www.freepress.net/our-response/expert-analysis/explainers/net-neutrality-violations-brief-history}.

\bibitem{khattak16ndss}
{\sc Khattak, S., Fifield, D., Afroz, S., Javed, M., Sundaresan, S., Paxson,
  V., Murdoch, S.~J., and McCoy, D.}
\newblock Do you see what {I} see? {D}ifferential treatment of anonymous users.
\newblock In {\em Proceedings of the Network and Distributed System Security
  Symposium (NDSS)\/} (2016).

\bibitem{khattak2013towards}
{\sc Khattak, S., Javed, M., Anderson, P.~D., and Paxson, V.}
\newblock Towards illuminating a censorship monitor's model to facilitate
  evasion.
\newblock In {\em The 3rd USENIX Workshop on Free and Open Communications on
  the Internet\/} (2013), USENIX.

\bibitem{lomas18tc}
{\sc Lomas, N.}
\newblock {WTF} is {GDPR}?
\newblock {\em TechCrunch\/} (Jan. 2018).
\newblock \url{https://techcrunch.com/2018/01/20/wtf-is-gdpr/}.

\bibitem{Lowe2007a}
{\sc Lowe, G., Winters, P., and Marcus, M.~L.}
\newblock The great {DNS} wall of {China}.
\newblock Tech. rep., New York University, 2007.

\bibitem{Marczak2015b}
{\sc Marczak, B., Weaver, N., Dalek, J., Ensafi, R., Fifield, D., McKune, S.,
  Rey, A., Scott-Railton, J., Deibert, R., and Paxson, V.}
\newblock An analysis of {China's} {``Great Cannon''}.
\newblock In {\em Free and Open Communications on the Internet (FOCI)\/}
  (2015), USENIX.

\bibitem{maurer14slate}
{\sc Maurer, T., and Morgus, R.}
\newblock Stop calling decentralization of the internet ``balkanization''.
\newblock {\em Slate\/} (Feb. 2014).
\newblock In the Future Tense blog:
  \url{http://www.slate.com/blogs/future_tense/2014/02/19/stop_calling_decentralization_of_the_internet_balkanization.html}.

\bibitem{mikians12hotnets}
{\sc Mikians, J., Gyarmati, L., Erramilli, V., and Laoutaris, N.}
\newblock Detecting price and search discrimination on the internet.
\newblock In {\em Proceedings of the 11th ACM Workshop on Hot Topics in
  Networks\/} (New York, NY, USA, 2012), ACM, pp.~79--84.

\bibitem{mikians13conext}
{\sc Mikians, J., Gyarmati, L., Erramilli, V., and Laoutaris, N.}
\newblock Crowd-assisted search for price discrimination in e-commerce: First
  results.
\newblock In {\em Proceedings of the Ninth ACM Conference on Emerging
  Networking Experiments and Technologies\/} (New York, NY, USA, 2013), ACM,
  pp.~1--6.

\bibitem{nabi2013anatomy}
{\sc Nabi, Z.}
\newblock The anatomy of web censorship in {Pakistan}.
\newblock In {\em Presented as part of the 3rd USENIX Workshop on Free and Open
  Communications on the Internet\/} (Berkeley, CA, 2013), USENIX.

\bibitem{park2010empirical}
{\sc Park, J.~C., and Crandall, J.~R.}
\newblock Empirical study of a national-scale distributed intrusion detection
  system: Backbone-level filtering of {HTML} responses in {China}.
\newblock In {\em Distributed Computing Systems\/} (2010), IEEE, pp.~315--326.

\bibitem{parrilla18gdpr}
{\sc Parrilla, D.}
\newblock {GDPR} for lazy people: {B}lock all {E}uropean users with
  {C}loudflare {W}orkers.
\newblock
  \url{https://apility.io/2018/05/25/gdpr-lazy-block-european-users-cloudflare-workers/},
  May 2018.

\bibitem{paukner13news}
{\sc Paukner, P.}
\newblock {D}iese {K}ultur ist in {D}eutschland leider nicht verfügbar.
\newblock {\em S\"{u}ddeutsche Zeitung\/} (2013).
\newblock
  \url{http://www.sueddeutsche.de/digital/streit-zwischen-youtube-und-gema-diese-kultur-ist-in-deutschland-leider-nicht-verfuegbar-1.1584813}.

\bibitem{sfakianakis2011censmon}
{\sc Sfakianakis, A., Athanasopoulos, E., and Ioannidis, S.}
\newblock {CensMon}: A web censorship monitor.
\newblock In {\em Free and Open Communications on the Internet\/} (2011),
  USENIX.

\bibitem{singh17usenix}
{\sc Singh, R., Nithyanand, R., Afroz, S., Pearce, P., Tschantz, M.~C., Gill,
  P., and Paxson, V.}
\newblock Characterizing the nature and dynamics of {T}or exit blocking.
\newblock In {\em {USENIX} Security\/} (Aug. 2017).

\bibitem{unblockvideosXXweb}
{\sc {UnblockVideos.com}}.
\newblock {YouTube} region restriction statistics, check {YouTube} video
  restrictions online.
\newblock \url{https://unblockvideos.com/youtube-video-restriction-checker/}.
\newblock Accessed May~28, 2018.

\bibitem{verkamp2012inferring}
{\sc Verkamp, J.-P., and Gupta, M.}
\newblock Inferring mechanics of web censorship around the world.
\newblock In {\em Free and Open Communications on the Internet\/} (2012),
  USENIX.

\bibitem{vissers14hotpets}
{\sc Vissers, T., Nikiforakis, N., Bielova, N., and Joosen, W.}
\newblock Crying wolf? {O}n the price discrimination of online airline tickets.
\newblock In {\em 7th Workshop on Hot Topics in Privacy Enhancing Technologies
  (HotPETs 2014)\/} (July 2014).

\bibitem{winters12foci}
{\sc Winter, P., and Lindskog, S.}
\newblock How the {Great Firewall of China} is blocking {Tor}.
\newblock In {\em Free and Open Communications on the Internet\/} (Bellevue,
  WA, USA, 2012), USENIX.

\bibitem{wired04}
{\sc {Wired Staff}}.
\newblock Solution for slashdot effect?
\newblock {\em Wired\/} (10 2004).
\newblock \url{https://www.wired.com/2004/10/solution-for-slashdot-effect/}.

\bibitem{xu2011internet}
{\sc Xu, X., Mao, Z.~M., and Halderman, J.~A.}
\newblock Internet censorship in {China}: Where does the filtering occur?
\newblock In {\em Passive and Active Measurement Conference\/} (Atlanta, GA,
  USA, 2011), Springer, pp.~133--142.

\bibitem{zheleva13dev}
{\sc Zheleva, M., Schmitt, P., Vigil, M., and Belding, E.}
\newblock The increased bandwidth fallacy: Performance and usage in rural
  zambia.
\newblock In {\em Proceedings of the 4th Annual Symposium on Computing for
  Development\/} (New York, NY, USA, 2013), ACM DEV-4 '13, ACM, pp.~2:1--2:10.

\bibitem{zittrain2003internet}
{\sc Zittrain, J., and Edelman, B.~G.}
\newblock Internet filtering in {China}.
\newblock {\em IEEE Internet Computing 7}, 2 (Mar. 2003), 70--77.

\end{thebibliography}
